\input harvmac
\input epsf

\let\includefigures=\iftrue
\let\useblackboard=\iftrue
\newfam\black


\includefigures
\message{If you do not have epsf.tex (to include figures),}
\message{change the option at the top of the tex file.}

\def\figin{\epsfcheck\figin}\def\figins{\epsfcheck\figins}
\def\epsfcheck{\ifx\epsfbox\UnDeFiNeD
\message{(NO epsf.tex, FIGURES WILL BE IGNORED)}
\gdef\figin##1{\vskip2in}\gdef\figins##1{\hskip.5in}
\else\message{(FIGURES WILL BE INCLUDED)}%
\gdef\figin##1{##1}\gdef\figins##1{##1}\fi}
\def\DefWarn#1{}
\def\figinsert{\goodbreak\midinsert}
\def\ifig#1#2#3{\DefWarn#1\xdef#1{fig.~\the\figno}
\writedef{#1\leftbracket fig.\noexpand~\the\figno}%
\figinsert\figin{\centerline{#3}}\medskip\centerline{\vbox{
\baselineskip12pt\advance\hsize by -1truein
\noindent\footnotefont{\bf Fig.~\the\figno:} #2}}
\endinsert\global\advance\figno by1}
\else
\def\ifig#1#2#3{\xdef#1{fig.~\the\figno}
\writedef{#1\leftbracket fig.\noexpand~\the\figno}%
\global\advance\figno by1} \fi

\def\id{{1 \kern-.28em {\rm l}}}

\def\K3{{\bf K3}}
\def\journal#1&#2(#3){\unskip, \sl #1\ \bf #2 \rm(19#3) }
\def\andjournal#1&#2(#3){\sl #1~\bf #2 \rm (19#3) }

\def\bar{\overline}

\def\frac#1#2{{#1\over#2}}

\def\inbar{\,\vrule height1.5ex width.4pt depth0pt}
\def\IC{\relax\hbox{$\inbar\kern-.3em{\rm C}$}}
\def\IR{\relax{\rm I\kern-.18em R}}
\def\IP{\relax{\rm I\kern-.18em P}}

%
%

%
\catcode`\@=11
\def\slash#1{\mathord{\mathpalette\c@ncel{#1}}}
\overfullrule=0pt

\def\underrel#1\over#2{\mathrel{\mathop{\kern\z@#1}\limits_{#2}}}

\catcode`\@=12


%



\lref\ChakrabortyMDF{
  S.~Chakraborty, A.~Giveon and D.~Kutasov,
  ``$T\bar{T}$, $J\bar{T}$, $T\bar{J}$ and String Theory,''
J.\ Phys.\ A {\bf 52}, no. 38, 384003 (2019).
[arXiv:1905.00051 [hep-th]].
}

\lref\GuicaLIA{
  M.~Guica,
  ``An integrable Lorentz-breaking deformation of two-dimensional CFTs,''
SciPost Phys.\  {\bf 5}, no. 5, 048 (2018).
[arXiv:1710.08415 [hep-th]].
}

\lref\ChakrabortyVJA{
  S.~Chakraborty, A.~Giveon and D.~Kutasov,
  ``$ J\overline{T} $ deformed CFT$_{2}$ and string theory,''
JHEP {\bf 1810}, 057 (2018).
[arXiv:1806.09667 [hep-th]].
}

\lref\ApoloQPQ{
  L.~Apolo and W.~Song,
  ``Strings on warped AdS$_{3}$ via $ T\bar{J} $ deformations,''
JHEP {\bf 1810}, 165 (2018).
[arXiv:1806.10127 [hep-th]].
}

\lref\HashimotoWCT{
  A.~Hashimoto and D.~Kutasov,
  ``$T \bar{T},J \bar T$, $T \bar{J}$ Partition Sums From String Theory,''
[arXiv:1907.07221 [hep-th]].
}

\lref\GiveonNS{
  A.~Giveon, D.~Kutasov and N.~Seiberg,
  ``Comments on string theory on AdS(3),''
Adv.\ Theor.\ Math.\ Phys.\  {\bf 2}, 733 (1998).
[hep-th/9806194].
}

\lref\KutasovXU{
  D.~Kutasov and N.~Seiberg,
  ``More comments on string theory on AdS(3),''
JHEP {\bf 9904}, 008 (1999).
[hep-th/9903219].
}

\lref\AraujoRHO{
  T.~Araujo, E.~Ó.~Colgáin, Y.~Sakatani, M.~M.~Sheikh-Jabbari and H.~Yavartanoo,
  ``Holographic integration of $T \bar{T}$ \& $J \bar{T}$ via $O(d,d)$,''
JHEP {\bf 1903}, 168 (2019).
[arXiv:1811.03050 [hep-th]].
}

\lref\GiveonNIE{
  A.~Giveon, N.~Itzhaki and D.~Kutasov,
  ``$ {T}\overline{{T}} $ and LST,''
JHEP {\bf 1707}, 122 (2017).
[arXiv:1701.05576 [hep-th]].
}

\lref\ArgurioTB{
  R.~Argurio, A.~Giveon and A.~Shomer,
  ``Superstrings on AdS(3) and symmetric products,''
JHEP {\bf 0012}, 003 (2000).
[hep-th/0009242].
}

\lref\NishiokaUN{
  T.~Nishioka, S.~Ryu and T.~Takayanagi,
  ``Holographic Entanglement Entropy: An Overview,''
J.\ Phys.\ A {\bf 42}, 504008 (2009).
[arXiv:0905.0932 [hep-th]].
}

\lref\KlebanovWS{
  I.~R.~Klebanov, D.~Kutasov and A.~Murugan,
  ``Entanglement as a probe of confinement,''
Nucl.\ Phys.\ B {\bf 796}, 274 (2008).
[arXiv:0709.2140 [hep-th]].
}

\lref\HolzheyWE{
  C.~Holzhey, F.~Larsen and F.~Wilczek,
  ``Geometric and renormalized entropy in conformal field theory,''
Nucl.\ Phys.\ B {\bf 424}, 443 (1994).
[hep-th/9403108].
}

\lref\RyuBV{
  S.~Ryu and T.~Takayanagi,
  ``Holographic derivation of entanglement entropy from AdS/CFT,''
Phys.\ Rev.\ Lett.\  {\bf 96}, 181602 (2006).
[hep-th/0603001].
}

\lref\CalabreseQY{
  P.~Calabrese and J.~Cardy,
  ``Entanglement entropy and conformal field theory,''
J.\ Phys.\ A {\bf 42}, 504005 (2009).
[arXiv:0905.4013 [cond-mat.stat-mech]].
}

\lref\ChakrabortyKPR{
  S.~Chakraborty, A.~Giveon, N.~Itzhaki and D.~Kutasov,
  ``Entanglement beyond AdS,''
Nucl.\ Phys.\ B {\bf 935}, 290 (2018).
[arXiv:1805.06286 [hep-th]].
}

\lref\CasiniES{
  H.~Casini and M.~Huerta,
  ``A c-theorem for the entanglement entropy,''
J.\ Phys.\ A {\bf 40}, 7031 (2007).
[cond-mat/0610375].
}

\lref\BF{
P.~F.~Byrd and M.~D.~Friedman,
``Handbook of Elliptic Integrals for Engineers and Scientists,"
 Springer Verlag, New York 1971.
}

\lref\CasiniHU{
  H.~Casini and M.~Huerta,
  ``Universal terms for the entanglement entropy in 2+1 dimensions,''
Nucl.\ Phys.\ B {\bf 764}, 183 (2007).
[hep-th/0606256].
}

\lref\GrieningerZTS{
  S.~Grieninger,
  ``Entanglement entropy and $T\bar T$ deformations beyond antipodal points from holography,''
[arXiv:1908.10372 [hep-th]].
}

\lref\SmirnovLQW{
  F.~A.~Smirnov and A.~B.~Zamolodchikov,
  ``On space of integrable quantum field theories,''
Nucl.\ Phys.\ B {\bf 915}, 363 (2017).
[arXiv:1608.05499 [hep-th]].
}

\lref\CavagliaODA{
  A.~Cavaglià, S.~Negro, I.~M.~Szécsényi and R.~Tateo,
  ``$T \bar{T}$-deformed 2D Quantum Field Theories,''
JHEP {\bf 1610}, 112 (2016).
[arXiv:1608.05534 [hep-th]].
}

\lref\GiveonMYJ{
  A.~Giveon, N.~Itzhaki and D.~Kutasov,
  ``A solvable irrelevant deformation of AdS$_{3}$/CFT$_{2}$,''
JHEP {\bf 1712}, 155 (2017).
[arXiv:1707.05800 [hep-th]].
}

\lref\AharonyBAD{
  O.~Aharony, S.~Datta, A.~Giveon, Y.~Jiang and D.~Kutasov,
  ``Modular invariance and uniqueness of $T\bar{T}$ deformed CFT,''
JHEP {\bf 1901}, 086 (2019).
[arXiv:1808.02492 [hep-th]].
}

\lref\AharonyICS{
  O.~Aharony, S.~Datta, A.~Giveon, Y.~Jiang and D.~Kutasov,
  ``Modular covariance and uniqueness of $J\bar{T}$ deformed CFTs,''
JHEP {\bf 1901}, 085 (2019).
[arXiv:1808.08978 [hep-th]].
}

\lref\MaldacenaUZ{
  J.~M.~Maldacena, J.~Michelson and A.~Strominger,
  ``Anti-de Sitter fragmentation,''
JHEP {\bf 9902}, 011 (1999).
[hep-th/9812073].
}

\lref\SeibergXZ{
  N.~Seiberg and E.~Witten,
  ``The D1 / D5 system and singular CFT,''
JHEP {\bf 9904}, 017 (1999).
[hep-th/9903224].
}

\lref\DattaTHY{
  S.~Datta and Y.~Jiang,
  ``$T\bar{T}$ deformed partition functions,''
JHEP {\bf 1808}, 106 (2018).
[arXiv:1806.07426 [hep-th]].
}

\lref\MaldacenaHW{
  J.~M.~Maldacena and H.~Ooguri,
  ``Strings in AdS(3) and SL(2,R) WZW model 1.: The Spectrum,''
J.\ Math.\ Phys.\  {\bf 42}, 2929 (2001).
[hep-th/0001053].
}

\lref\KutasovXU{
  D.~Kutasov and N.~Seiberg,
  ``More comments on string theory on AdS(3),''
JHEP {\bf 9904}, 008 (1999).
[hep-th/9903219].
}

\lref\BarbonUT{
  J.~L.~F.~Barbon and C.~A.~Fuertes,
  ``Holographic entanglement entropy probes (non)locality,''
JHEP {\bf 0804}, 096 (2008).
[arXiv:0803.1928 [hep-th]].
}

\lref\ShibaJJA{
  N.~Shiba and T.~Takayanagi,
  ``Volume Law for the Entanglement Entropy in Non-local QFTs,''
JHEP {\bf 1402}, 033 (2014).
[arXiv:1311.1643 [hep-th]].
}

\lref\GiveonNIE{
  A.~Giveon, N.~Itzhaki and D.~Kutasov,
  ``$ {T}\overline{{T}} $ and LST,''
JHEP {\bf 1707}, 122 (2017).
[arXiv:1701.05576 [hep-th]].
}

\lref\RyuEF{
  S.~Ryu and T.~Takayanagi,
  ``Aspects of Holographic Entanglement Entropy,''
JHEP {\bf 0608}, 045 (2006).
[hep-th/0605073].
}

\lref\RyuBV{
  S.~Ryu and T.~Takayanagi,
  ``Holographic derivation of entanglement entropy from AdS/CFT,''
Phys.\ Rev.\ Lett.\  {\bf 96}, 181602 (2006).
[hep-th/0603001].
}

\lref\McGoughLOL{
  L.~McGough, M.~Mezei and H.~Verlinde,
  ``Moving the CFT into the bulk with $ T\overline{T} $,''
JHEP {\bf 1804}, 010 (2018).
[arXiv:1611.03470 [hep-th]].
}

\lref\BzowskiPCY{
  A.~Bzowski and M.~Guica,
  ``The holographic interpretation of $J \bar T$-deformed CFTs,''
JHEP {\bf 1901}, 198 (2019).
[arXiv:1803.09753 [hep-th]].
}

\lref\LewkowyczXSE{
  A.~Lewkowycz, J.~Liu, E.~Silverstein and G.~Torroba,
  ``$ T\overline{T} $ and EE, with implications for (A)dS subregion encodings,''
JHEP {\bf 2004}, 152 (2020).
[arXiv:1909.13808 [hep-th]].
}

\lref\LewkowyczNQA{
  A.~Lewkowycz and J.~Maldacena,
  ``Generalized gravitational entropy,''
JHEP {\bf 1308}, 090 (2013).
[arXiv:1304.4926 [hep-th]].
}

\lref\HubenyXT{
  V.~E.~Hubeny, M.~Rangamani and T.~Takayanagi,
  ``A Covariant holographic entanglement entropy proposal,''
JHEP {\bf 0707}, 062 (2007).
[arXiv:0705.0016 [hep-th]].
}

\lref\ZamolodchikovCE{
  A.~B.~Zamolodchikov,
  ``Expectation value of composite field T anti-T in two-dimensional quantum field theory,''
[hep-th/0401146].
}

\lref\AsratJSH{
  M.~Asrat,
  ``KdV Charges and the Generalized Torus Partition Sum in $T{\bar T}$ deformation,''
Nucl.\ Phys.\ B {\bf 958}, 115119 (2020).
[arXiv:2002.04824 [hep-th]].
}


\Title{
} {\vbox{ \centerline{Entropic c--functions in $T{\bar T}, J{\bar T}, T{\bar J}$ deformations} }}

\bigskip
\centerline{\it Meseret Asrat}
\bigskip
\smallskip
\centerline{Enrico Fermi Institute and Department of Physics}
\centerline{University of Chicago} 
\centerline{5640 S. Ellis Av., Chicago, IL 60637, USA }

\smallskip

\vglue .3cm

\bigskip

\let\includefigures=\iftrue
\bigskip
\noindent

We study the holographic entanglement entropy of an interval in a quantum field theory obtained by deforming a holographic two--dimensional conformal field theory via a general linear combination of irrelevant operators that are closely related to, but nonetheless distinct from, $T{\bar T}, \ J{\bar T}$ and $T{\bar J}$, and compute the Casin--Huerta entropic $c$--function. We find that the entropic $c$--function is ultraviolet regulator independent, and along the renormalization group upflow towards the ultraviolet, it is non--decreasing. We show that the entropic $c$--function exhibits a power law divergence as the interval length approaches a minimum finite value determined in terms of the deformation parameters. We also find for a particular combination of the deformation parameters a square root correction of the entanglement entropy area law.

\bigskip

\Date{11/19}


\newsec{Introduction}

String theory on the background $AdS_3 \times {\cal X}$ with Neveu--Schwartz two--form $B$ field contains a sector of long strings \refs{\MaldacenaHW, \ \MaldacenaUZ, \ \SeibergXZ, \ \ArgurioTB}. The effective theory of $N$ coincident long strings on $AdS_3$ in a wide range of positions in the radial direction are believed to be well described by the symmetric product theory \refs{\ArgurioTB, \ \ChakrabortyMDF }
\eqn\symP{{{\cal M}^N/ S_N},
}
where the conformal field theory ${\cal M}$ is the theory of a single long string that extends to the boundary \SeibergXZ. Note that the symmetric product theory \symP\ is not equal to the full dual boundary (spacetime) theory since ${\cal M}$ only describes the long string sector. 

In \ChakrabortyMDF\ the authors considered deforming the Lagrangian of the worldsheet string theory on $AdS_3 \times {\cal X}$ by a general linear combination of truly marginal worldsheet current--current operators 
\eqn\ccdeff{\lambda J^-_{\rm SL} {\bar J}^-_{\rm SL} + \epsilon_+ K {\bar J}^-_{\rm SL} + \epsilon_- {\bar K} J^-_{\rm SL}.
}
They showed that, in the long string sector, this deformation is equivalent to the deformation of the theory ${\cal M}$ (with spectral flow or winding number $w = 1$) by a general linear combination of the recently much studied irrelevant operators \refs{\ZamolodchikovCE, \ \CavagliaODA, \ \SmirnovLQW, \ \GuicaLIA}
\eqn\irrdeff{-tT{\bar T} - \mu_+ J{\bar T} - \mu_ -{\bar J}T,
}
where $J$ and ${\bar J}$ are the left and right moving  $U(1)$ currents, respectively, and $T$ and ${\bar T}$ are the left and right moving stress tensor components, respectively, of the conformal field theory ${\cal M}$. 

String theory on $AdS_3 \times {\cal X}$ contains a class of vertex operators ${\bar A}, \ A$ and $D$ constructed in \KutasovXU. In the boundary (spacetime) theory the operators ${\bar A}, \ A$ and $D$ have the same scaling dimensions as $T{\bar J}, \ J{\bar T}$ and $T{\bar T}$, respectively, however, they are single trace operators. It is shown in \refs{\ChakrabortyMDF, \ \GiveonNIE, \ \ApoloQPQ} that the deformation \ccdeff\ or equivalently, in the long string sector, the deformation \irrdeff\ amounts to deforming the dual full boundary (spacetime) conformal field theory by a linear combination of the operators ${\bar A}, \ A$ and $D$. In particular, it is shown in \ChakrabortyMDF\ that, in the long string sector, the operators ${\bar A}, \ A$ and $D$ are proportional to the sum over the $N$ copies of the operators $T{\bar J}, \ J{\bar T}$ and $T{\bar T}$, respectively, of the field theory ${\cal M}$. 

The operators $J{\bar T}$ and $T{\bar J}$ have left and right scaling dimensions $(1, 2)$ and $(2, 1)$, respectively, and therefore, the deformation \irrdeff \ results in a theory that breaks Lorentz invariance \refs{\GuicaLIA, \ \ChakrabortyVJA, \ \ApoloQPQ}.

The spacetime couplings $t, \ \mu_+$ and $\mu_-$ are related to the worldsheet dimensionless couplings $\lambda,\ \epsilon_+$ and $\epsilon_-$ via the relations \ChakrabortyMDF,
\eqn\coupling{t = \pi \alpha' \lambda, \quad \mu_{\pm} = 2\sqrt{2\alpha'}\epsilon_{\pm}, \quad \alpha' = l_s^2,
}
where $\alpha'$ is the Regge slope, and $l_s$ is the intrinsic string length. 

The deformation \irrdeff \ is irrelevant and therefore the couplings grow as we ascend the renormalization group. Under irrelevant deformation of a quantum field theory, in general, in the ultraviolet, we are in the strong coupling regime and the description of the theory in terms of the original infrared degrees of freedom often breaks down. The theory also suffers from ambiguities and/or arbitrariness, and quantum corrections generate infinite number of irrelevant operators. However, it is shown that under the deformation \irrdeff \ the theory is unambiguous (at least for one sign of the coupling $t$) and it is solvable at finite values of the couplings, in the sense that the spectrum on an infinite cylinder \refs{\CavagliaODA, \ \SmirnovLQW, \ \GuicaLIA} and the partition function on a torus \refs{\DattaTHY, \ \AharonyBAD, \ \AharonyICS, \ \HashimotoWCT, \ \AsratJSH} can be computed exactly. We also obtain in this paper the exact entropic c--function. This increases the number of quantities that one can exactly solve and study in this class of theories.

It is shown in \ChakrabortyMDF\ that in the space of couplings in which the combination
\eqn\master{F =  {t\over \pi} - {(\mu_+ + \mu_-)^2\over 8},
}
is positive, $F > 0$, the energies of states are real, and the density of states asymptotically exhibits Hagedorn growth. In the limit $F \to 0^+$ however the theory appears to be distinct. The density of states (in a fixed charge sector) asymptotically exhibits an intermediate growth between Cardy and Hagedorn growths. We also show later in the paper that the Von Neumann entanglement entropy in this limit exhibits a square root area law correction at next--to--leading order at short distances. In the case in which $F < 0$ the energies become complex above a scale fixed by the couplings and the corresponding bulk geometry is singular and/or it has either closed timelike curves or no timlike direction. The signature of the bulk metric switches signs beyond a finite radial distance where the singularity occurs. We will not consider this case in this paper as it is not clear how to consistently apply the Ryu and Takayanagi holographic prescription and its covariant generalization. We comment on this later in the discussion section.

We now briefly mention the holographic proposals for the closely related double trace deformations. In this class of deformations $T$ and $J$ are the left moving energy momentum tensor and current, respectively, of the full boundary theory. The right moving energy momentum tensor and current are denoted by $\bar T$ and $\bar J$ respectively. In these holographic proposals the coupling $t$ is negative and therefore $F < 0$. For negative $t$ with $\mu_\pm = 0$ the dual bulk spacetime of a $T{\bar T}$ deformed two dimensional holographic conformal field theory is proposed to be $AdS_3$ with a Dirichlet boundary at finite radial distance fixed by the coupling $t$ \McGoughLOL. For either sign of $\mu_+$ (and $\mu_- = 0$ or vice versa) and $t = 0$ it is shown in \BzowskiPCY\ that the dual bulk spacetime of a $J{\bar T}$ (or $T{\bar J}$) deformed two dimensional holographic conformal field theory is $AdS_3$ with boundary conditions that mix the metric and a gauge field dual to the current $J$ (or $\bar J$). In either of these cases in the field theory side we have states with complex energies.

In this paper we study the (Von Neumann) entanglement entropy for a spatial interval of length $L$ in the deformed (full) spacetime theory with $F \geq 0$ from its bulk string theory description and compute the Casin--Huerta entropic $c$--function. We study the monotonicity property of the c--function along the renormalization group upflow and its independence of regularization scheme that one often introduces to regularize the ultraviolet divergence of entanglement entropy. This provides a non--trivial consistency check of the deformed theory. It also gives insight into the nature of the theory in the ultraviolet as it is not governed by an ultraviolet fixed point. 

The rest of the paper is organized as follows. In section 2 we review the corresponding bulk string theory background obtained under the deformation \ccdeff. In section 3 we compute the entanglement entropy in the deformed dual spacetime theory using the holographic prescription. Following this, we discuss its large and small $L$ limits. In section 4 we compute the entropic $c$--function and study its ultraviolet and infrared limits. In section 5 we discuss the main results and future research directions. We comment on the entropic $c$--function of a double trace deformed field theory. We also comment on the case where $F < 0$. In the appendix we collect some of the intermediate results that are required in section 3.

\ifig\figc{Entropic c--function $C$ per $c_0$ as a function of interval length $L$ per $L_0$. Where $c_0 = {c\over 3}$ and $L_0 = {1\over 2}\sqrt{\pi t}$. The orange plot is for $\delta = 0$, and the blue plot is for $\delta^2 = 1$. The (normalized) entropic c--function for the case $\delta = 0$ diverges at $L_0$.}
{\epsfxsize3.6in\epsfbox{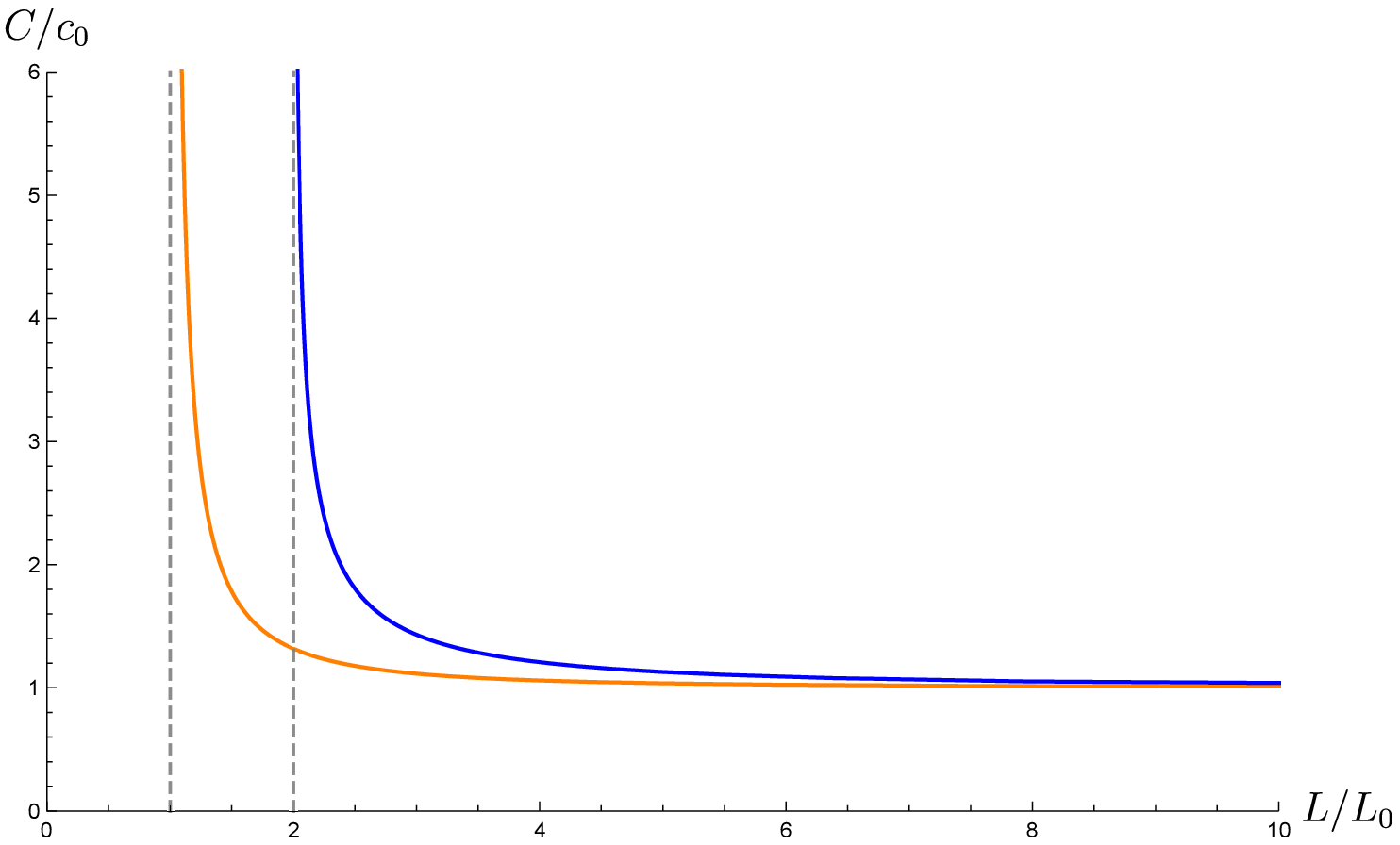}}

\newsec{The bulk deformed string background}

We begin with string theory on 
\eqn\back{AdS_3 \times S^1 \times {\cal N},
} with Neveu--Schwartz two--form $B$ field. Where the component ${\cal N}$ is an internal six--dimensional compact manifold. Its presence is irrelevant in our discussion. The $S^1$ component gives rise in the boundary conformal field theory to a $U(1)$ current algebra generated by the spacetime currents $J$ and ${\bar J}$ \refs{\GiveonNS, \ \KutasovXU}. 

The bosonic part of the worldsheet theory on $AdS_3 \times S^1$ is described by the action
\eqn\adss{S = {k\over 2\pi}\int d^2z(\partial \phi{\bar \partial}\phi  + e^{2\phi} \partial {\bar \gamma} {\bar \partial}\gamma + {1\over k}\partial \psi {\bar \partial \psi}),
} 
where $(\phi, \gamma, {\bar \gamma})$ are the coordinates on $AdS_3$, and $\psi \sim \psi + 2\pi$ is the coordinate on $S^1$. The boundary of $AdS_3$ is located at $\phi = +\infty$. The coordinates $\gamma$ and $\bar \gamma$ are
\eqn\space{l_s\gamma = t + x, \quad l_s \bar \gamma = -t + x.
}
The level $k$ is given by
\eqn\level{l^2 = l_s^2k,
}
where $l$ is the radius of curvature of $AdS_3$.

The action has an affine $SL(2, R)_{ L} \times SL(2, R)_{ R} \times U(1)_{ L} \times U(1)_{ R}$ symmetry with left mover worldsheet currents  
\eqn\curr{J^-_{\rm SL} = e^{2\phi}\partial\bar\gamma, \quad J^+_{\rm SL} = -2\gamma \partial\phi - \partial\gamma + \gamma^2 e^{2\phi}\partial\bar\gamma, \quad J^3_{\rm SL} = \gamma e^{2\phi}\partial\bar\gamma - \partial \phi, \quad K = \partial \psi,
}
and similar expressions for the right movers ${\bar J}^-_{\rm SL}, {\bar J}^+_{\rm SL}$, ${\bar J}^3_{\rm SL}$ and ${\bar K}$.

Consider deforming the worldsheet theory \adss\ by adding to its Lagrangian the deformation \ccdeff. The deformation \ccdeff\ is truly marginal, and therefore, it preserves the conformal symmetry. It breaks the affine $SL(2, R)_{L} \times SL(2, R)_{R} \times U(1)_{L} \times U(1)_{R}$ worldsheet symmetry down to $U(1)_{L}\times U(1)_{R} \times U(1)_{L} \times U(1)_{R}$ affine symmetry. 

The deformation corresponds to a deformation of the metric $g$, dilaton $\Phi$, and Neveu--Schwartz two--form $B$ \refs{\ChakrabortyMDF, \ \AraujoRHO}. We shall refer to the deformed background as ${\cal M}_4$,
\eqn\metric{ds^2 = d\phi^2 + hd\gamma d{\bar\gamma} + {2h\epsilon_+\over \sqrt{k}}d\psi d{\bar\gamma} + {2h\epsilon_-\over \sqrt{k}}d\psi d\gamma + {1\over k}hf^{-1}d\psi^2,
}
\eqn\dilaton{e^{2\Phi} = g_s^2e^{-2\phi}h,
}
\eqn\bfield{B_{\gamma{\bar\gamma}} = g_{\gamma{\bar\gamma}}, \quad B_{\gamma\psi} = g_{\gamma\psi}, \quad B_{\psi {\bar\gamma}} = g_{{\bar \gamma}\psi},
}
where
\eqn\paradef{h^{-1} = e^{-2\phi} + \lambda - 4\epsilon_+\epsilon_-, \quad f^{-1}= h^{-1} + 4\epsilon_+\epsilon_-.
}

For $\lambda = 0,\ \epsilon_\pm = 0$ ${\cal M}_4$ reduces to our starting background $AdS_3 \times S^1$. For $\lambda = 0$ ${\cal M}_4$ reduces to a warped $AdS_3 \times S^1$ background \refs{\ChakrabortyVJA, \ \ApoloQPQ}. For $\epsilon_\pm  = 0$ ${\cal M}_4$ reduces to a background that is asymptotically $AdS_3 \times S^1$ for large negative $\phi$ and $\IR_\phi \times \IR^{1, 1}\times S^1$ for large positive $\phi$ \GiveonNIE. 

It is shown in \ChakrabortyMDF\ that for a combination of the couplings
\eqn\smooth{\Psi = \lambda - (\epsilon_+ + \epsilon_-)^2, 
}
with $\Psi \geq 0$ the geometry is smooth and it has no closed timelike curves. This positivity condition on \smooth\ is the dual analogue of the positivity condition on \master.

In this paper we consider the case in which
\eqn\cond{\epsilon_+ = {\epsilon\over 2}, \quad \epsilon_- = {\epsilon\over 2}, \quad \Psi \geq 0.
}
In this case the background ${\cal M}_4$ is
\eqn\metricss{ds^2 = \alpha'd\phi^2 - hdt^2 + h\left(dx + {\epsilon \sqrt{\alpha' \over k}}d\psi\right)^2 + {\alpha'\over k}d\psi^2, \quad e^{2\Phi} = g_s^2e^{-2\phi}h,  \quad h^{-1} = e^{-2\phi} + \Psi.
}
In what follows we work on this background to compute the entanglement entropy and the entropic $c$--function for an interval of length $L$ in the deformed dual conformal field theory using the holographic prescription. We note that for $\lambda < 0$ or equivalently $\Psi < 0$ the function $h$ changes sign beyond a finite radial distance and therefore the metric signature also changes signs. It is not clear in this case how to consistently apply the holographic prescription. We will not consider this case in this paper. We comment on this later in the discussion section.

\newsec{Holographic entanglement entropy}

In this section we compute the entanglement entropy in the deformed spacetime conformal field theory for a spatial interval of length $L$ with endpoints at $x = -L/2$ and $x = +L/2$.  It is defined as the Von Neumann entropy,
\eqn\VNum{
S = -{\rm Tr} \rho \log \rho,
}
corresponding to reduced density matrix $\rho$ of the subsystem $L$. The reduced density matrix $\rho$ is obtained by tracing the density matrix of the full system over the states of the complement of the subsystem $L$. The entanglement entropy measures the degree of entanglement between the subsystem $L$ and its complement. In holographic field theories, entanglement entropy is encoded in certain geometrical quantities in the bulk geometry. In what follows we begin by briefly stating the holographic entanglement prescription to compute entanglement entropy in quantum field theories that have dual string theory descriptions \refs{\RyuEF, \ \RyuBV, \ \NishiokaUN, \ \KlebanovWS, \ \LewkowyczNQA}. 

Suppose we have a $d$--dimensional holographic quantum field theory. Suppose also the dual string theory is on a background ${\cal M}_{d + 1}$. To compute the entanglement entropy  of a given spatial region ${\cal R}$ in the boundary quantum field theory we first find a co--dimension two static surface ${\cal K} $ in the bulk geometry ${\cal M}_{d+1}$ that ends on the boundary of ${\cal R}$ and minimizes the area functional. Then the entanglement entropy $S_{\cal R}$ in the $d$--dimensional boundary quantum field theory, in the large central charge limit, is given by
\eqn\entro{S_{\cal R} = {{\rm Area}({\cal K}) \over 4G_{\rm N}^{(d + 1)}},
}
where $G^{(d + 1)}_{\rm N}$ is the $d + 1$--dimensional Newton's constant of the ${\cal M}_{d + 1}$ geometry.

Following the holographic prescription we now compute the entanglement entropy for the spatial interval of length $L$ in the deformation dual to \ccdeff. The bulk string geometry \metricss\ at a moment of time is  
\eqn\metricsss{ds^2 = \alpha' d\phi^2 + hdy^2 + {\alpha'\over k}d\psi^2, \quad h^{-1} = e^{-2\phi} + \Psi,
}
where $y = x + {\epsilon\sqrt{\alpha'/k}} \ \psi$.

We now look for a two--dimensional surface $\phi(y, \psi)$ in the geometry \metricsss\ (and wrapping the internal ${\cal N}$ space) that minimizes globally (in the space of functions) the area functional \entro\ which taking into account the dilaton \refs{\RyuEF, \ \KlebanovWS} is\foot{Here we rescaled the metric \metricsss\ by the level $k$.}
\eqn\enTTTT{S = {\sqrt{k\alpha'}\over 4G_N^{(4)}}\int_0^{2\pi} d\psi \int_{-{L\over 2} + {\epsilon\sqrt{\alpha'/k}} \ \psi}^{+{L\over 2} + {\epsilon\sqrt{\alpha'/k}} \ \psi} dy \ e^{2\phi}\sqrt{{1\over h}\left(1 + {\alpha'\over h} \left(\partial_y\phi\right)^2+ k\left(\partial_\psi\phi\right)^2\right)}, 
}
with the boundary conditions
\eqn\bcs{\phi(\pm L/2 + {\epsilon \sqrt{\alpha'/k}} \ \psi, \psi) = \infty, \quad \phi(y, 0) = \phi(y + {\epsilon \sqrt{\alpha'/k}} \ 2\pi, 2\pi),
}
where $\psi$ is on $S^1$ that is $\psi \sim \psi + 2\pi$.

We note that under the following continuous and discrete spacetime transformations
\eqn\sym{\psi \to \psi + \delta, \quad y \to y + {\epsilon \sqrt{\alpha' \over k}}\delta, \quad {\rm and} \quad \psi \to -\psi + 2\pi, \quad y \to -y + {\epsilon \sqrt{\alpha' \over k}} 2\pi,
}
where $\delta$ is an arbitrary constant, the bulk background \metricsss\ and the boundary conditions \bcs\ are invariant. The surface 
\eqn\sol{\phi(y, \psi) = \phi(y - {\epsilon\sqrt{\alpha'/k}} \ \psi) = \phi(-y + {\epsilon\sqrt{\alpha'/k}} \ \psi),
}
is invariant under the above symmetry transformations \sym\ and thus we expect that it minimizes the area functional. The minimal surface \sol\ is generated by translating the curve, for example at $\psi = 0$, $\phi(y)$, along the line $y = {\epsilon\sqrt{\alpha'/k}} \ \psi$. This curve has the parity symmetry $y \to - y$. The entanglement entropy is then obtained using \enTTTT. We find 
\eqn\hoentr{ S = {\sqrt{k}\over 4 G_{N}^{(3)}}\int_{-{L\over 2}}^{+{L\over 2}} dx\sqrt{H(U)}\sqrt{1 + \beta(U)(\partial_xU)^2},
}
where $G^{(3)}_{\rm N} =  G^{(4)}_{\rm N} / 2\pi l_s$, and 
\eqn\deff{U = e^{\phi}, \quad U^2h^{-1} = 1 + U^2\left(\lambda - \epsilon^2\right) , \quad U^{-2}H(U) = U^2 h^{-1} , \quad U^4\beta(U) =  (1 + U^2\lambda)\alpha' .
}
The boundary conditions \bcs\ now take the form 
\eqn\bcss{U(\pm L/2) = U_{\infty},
} 
where $U_{\infty}$ is an ultraviolet cutoff.

We denote the value at which the curve $U$ takes its minimum value by $U_0$. This value is related to the length of the interval $L$.  This follows from the Euler's variational equation of the action \hoentr\ with the boundary conditions \bcss. One finds
\eqn\intt{L(U_0) = 2\sqrt{H(U_0)}\int_{U_0}^{U_\infty} dU{\sqrt{\beta(U)}\over\sqrt{H(U) - H(U_0)}}.
}
The entropy using the Euler's equation of motion becomes 
\eqn\entr{S = {\sqrt{k}\over  2G_{\rm N}^{(3)}}\int_{U_0}^{U_\infty} dU \sqrt{\beta(U)\over H(U) - H(U_0)}H(U).
}

We rewrite the expression \intt\ of the interval length $L$ in terms of the minimum value $U_0$ as
\eqn\strlen{L(U_0) =  {\sqrt{\alpha'}\over U_0} \int_1^{x_\infty}{dx\over x}\sqrt{{(1 + \alpha x)(1 + \alpha_-) \over x(x - 1)( \alpha_- x + \alpha_- + 1) }} , 
}
where
\eqn\mdef {\alpha = \lambda U^2_0, \quad \alpha_- = \Psi U^2_0, \quad x_\infty = {U_\infty^2\over U_0^2}.
}

The integral \strlen\ is ultraviolet convergent and it solves to 
\eqn\sstrlen{{L \over 2\sqrt{\alpha'\lambda}}= \sqrt{1 + \alpha \over \alpha} E\left(\arcsin\sqrt{{1 + \alpha_- \over 1 + 2\alpha_-}}, \sqrt{1 + 2\alpha_- \over (1 + \alpha)(1 + \alpha_-)}\right),
}
where $E(\varphi, k)$ is the incomplete elliptic integral of the second kind,
\eqn\ellpti{E(\varphi, k) = \int_0^{\varphi} d\theta \sqrt{1 - k^2 \sin^2\theta}.
}

In the limit in which $\Psi \to 0^+$ or, equivalently $\alpha_- \to 0^+$, we note that the interval $L$ \sstrlen\ takes the following simpler form
\eqn\sstrlens{{L \over 2\sqrt{\alpha'\lambda}}= \sqrt{1 + \alpha \over \alpha} E\left(\sqrt{1 \over 1 + \alpha}\right),
}
where $E(k) = E(\pi/2, k)$ is the complete elliptic integral of the second kind. Sending $\alpha \to 0^+$ in \sstrlens\ yields
\eqn\intads{{L \over \sqrt{\alpha'}}= {2\over U_0}.
}

We rewrite the entanglement entropy \entr\ as
\eqn\enttr{S = {\sqrt{k\alpha'}\over  4G_{\rm N}^{(3)}}\int_1^{x_\infty}dx\sqrt{{ \alpha x + 1 \over x(x - 1)( \alpha_-x + \alpha_- + 1)  }} \cdot (\alpha_-x + 1).
}
We note that for $\alpha_- \neq 0$ the entropy diverges as $S \sim x_\infty$, and in the limit in which we take $\alpha_- \to 0^+$ with $\alpha \neq 0$ it diverges as $S \sim \sqrt{x_\infty}$. We also note that in the case in which we take both $\alpha \to 0^+, \ \alpha_- \to 0^+$ it diverges as $S \sim {\rm log}(x_\infty)$. The integral \enttr\ solves with the ultraviolet cutoff $x_\infty$ to 
\eqn\entros{ \eqalign{S & = {\sqrt{k\alpha'}\over  2G_{\rm N}^{(3)}} {1\over\sqrt{ (\alpha + 1)(\alpha_- + 1)}}\left\{ \left({ \alpha_-} + { \alpha}  - \alpha\alpha_-{d\over d\xi} \right)\right.\cr
&\left.\left[{1 \over \xi + 1} \cdot \Pi \left(\arcsin\sqrt{{\alpha_- + 1\over 2\alpha_- + 1}\cdot \left( {1-{1\over x_\infty}}\right)}, {2\alpha_- + 1 \over (\xi + 1)( \alpha_- + 1)}, \sqrt{{2\alpha_- + 1 \over (\alpha + 1)( \alpha_- + 1)}}\right)\right]_{\xi = 0} \right.\cr 
& \left. + \ F\left(\arcsin\sqrt{{\alpha_- + 1\over 2\alpha_- + 1}\cdot \left(1 - {1\over x_\infty}\right) }, \sqrt{{2\alpha_- + 1 \over (\alpha + 1)( \alpha_- + 1)}}\right)\right\},
}}
where $\Pi(\varphi, n, k)$ is the incomplete elliptic integral of the third kind, and $F(\varphi, k) = \Pi(\varphi, 0, k)$ is the incomplete elliptic integral of the first kind,
\eqn\elippi{\Pi(\varphi, n, k) = \int_0^{\varphi} {d\theta\over (1 - n\sin^2\theta)\sqrt{1 - k^2\sin^2\theta}}.
}

In the limit $\Psi \to 0^+$ or, equivalently $\alpha_- \to 0^+$, the entropy \entros\ gives
\eqn\enross{ \eqalign{S & = { \sqrt{k\alpha'}\sqrt{1 + \alpha }\over  2G_{\rm N}^{(3)}}\left[F\left(\arcsin\sqrt{ 1 - {1\over x_\infty} }, \sqrt{1\over 1 + \alpha}\right) - E\left(\arcsin\sqrt{ 1 - {1\over x_\infty} }, \sqrt{1\over 1 + \alpha}\right)\right] \cr
 &+  { \sqrt{k\alpha'}\over  2G_{\rm N}^{(3)}}\sqrt{(\alpha x_\infty + 1)\cdot \left(1 - {1\over x_\infty}\right) }.
}}
Taking $\alpha \to 0^+$ in \enross\ gives 
\eqn\entroads{S = { \sqrt{k\alpha'}\over  2G_{\rm N}^{(3)}}{\rm Log}(2\sqrt{x_\infty}). 
}

In the rest of the current section we study the above results in turns.

\subsec{Case $\Psi = 0: \lambda = 0, \ \epsilon = 0$}
In this case we have for the interval length $L$ and entanglement entropy $S$ from \intads\ and \entroads\
\eqn\adsel{{L \over \sqrt{\alpha'}}= {2\over U_0}, \quad S = { \sqrt{k\alpha'}\over  2G_{\rm N}^{(3)}}{\rm Log}\left(2{U_\infty\over U_0}\right).
}
We write the entropy as
\eqn\adsentr{ S = {c\over 3}{\rm Log}\left(2 {L\over L_{\Lambda}}\right), \quad {L_{\Lambda}\over \sqrt{\alpha'}} := {2\over U_\infty}, \quad c = { 3\sqrt{k\alpha'}\over  2G_{\rm N}^{(3)}},
}
where $L_{\Lambda}$ is an ultraviolet cutoff. This result is the well--known entanglement entropy for a two--dimensional holographic conformal field theory with (Brown--Henneaux) central charge $c$ that is dual to pure $AdS_3$ \refs{\HolzheyWE, \ \RyuBV, \ \CalabreseQY}.

\subsec{Case $\Psi = 0: \lambda = \epsilon^2 \neq 0$} 
In this case we have from \sstrlens\ and \enross\ that the interval length $L$ and entanglement entropy $S$ are given by
\eqn\sstrlenss{{L \over 2\sqrt{\alpha'\lambda}}= \sqrt{1 + \alpha \over \alpha} E\left(\sqrt{1 \over 1 + \alpha}\right),
}
\eqn\enrosss{ \eqalign{S & = { \sqrt{k\alpha'}\sqrt{1 + \alpha }\over  2G_{\rm N}^{(3)}}\left[F\left(\arcsin\sqrt{ 1 - {1\over x_\infty} }, \sqrt{1\over 1 + \alpha}\right) - E\left(\arcsin\sqrt{ 1 - {1\over x_\infty} }, \sqrt{1\over 1 + \alpha}\right)\right] \cr
 &+  { \sqrt{k\alpha'}\over  2G_{\rm N}^{(3)}}\sqrt{(\alpha x_\infty + 1)\cdot \left(1 - {1\over x_\infty}\right) }, \quad \alpha = \lambda U_0^2, \quad x_\infty = {U_\infty^2\over U_0^2}.
}}
Note that this case corresponds taking $F = 0$ in \master.

We find from \sstrlenss\ that in the large $U_0$ limit the interval length $L$ asymptotes to a minimum value which we denote by $L_0$. It takes the value
\eqn\smalll{L_0 = \pi \sqrt{\alpha'\lambda} = \sqrt{\pi t}.
}
We find using \sstrlenss\ the following large $U_0$ expansion of the interval length $L$, 
\eqn\lenn{{L\over L_0} = 1 + {1\over 4}\cdot {1\over \alpha} - {3\over 64} \cdot {1\over \alpha^2} +{\cal O}\left({1\over \alpha^3}\right), \quad \alpha = \lambda U_0^2.
}
Inverting the above equation one finds 
\eqn\lennnnv{\alpha = {1\over 4\xi} -{3\over 16} +{\cal O}(\xi), \quad \xi = {L\over L_0} - 1.
}

The small $U_0$ expansion of the interval length $L$ is
\eqn\llns{{L\over\sqrt{\alpha'}} = {2\over U_0}\left(1 - {1\over 4}\alpha \ln \alpha + {1\over 32}\alpha^2\ln \alpha + {\cal O}(\alpha^3)  \right) , \quad \alpha = \lambda U_0^2.
}
The leading term corresponds to the deep $AdS_3$ geometry (see \adsel). Therefore, in the large $L$ limit the surface is deep inside the bulk. We note also that the correction starts at order $\lambda = \epsilon^2$. 

Inverting equation \llns\ we find
\eqn\llnss{\alpha = \left({2\over \pi}\cdot {L_0\over L}\right)^2\left[1 - \left({2\over \pi}\cdot {L_0\over L}\right)^2{\rm Log}\left({2\over \pi}\cdot {L_0\over L}\right)+{\cal O}\left({2\over \pi}\cdot {L_0\over L}\right)^4 \right].
}

We can similarly study the large and small $U_0$ or, equivalently, the large and small $L$ limits of the entanglement entropy \enrosss. One finds in the large interval length $L$ limit
\eqn\entcaaav{S = { c\over 3}\left[\sqrt{\alpha x_\infty} -{1\over 2}{\rm Log}(\alpha) -{\alpha\over 8}{\rm Log}(\alpha) +{\cal O}\left(\alpha^2\right)\right], \quad \alpha = \lambda U_0^2, \quad x_\infty = {U_\infty^2\over U_0^2},
}
which upon using \llnss\ gives
\eqn\entcaa{S = { c\over 3}\left[{L_0\over L_{\Lambda}} -{\rm Log}\left({2\over \pi} \cdot {L_0\over L}\right) + {1\over 4}\left( {2\over \pi} \cdot {L_0\over L}\right)^2{\rm Log}\left({2\over \pi} \cdot  {L_0\over L}\right) + {\cal O}\left({2\over \pi}\cdot {L_0\over L}\right)^4 \right], 
}
where
\eqn\varFFFRRR{L_{\Lambda} := {\pi\sqrt{\alpha'}\over U_\infty},
}
 and $L_{\Lambda}$ is an ultraviolet cutoff. The leading logarithmic term is the contribution from the $AdS_3$ region found deep inside the bulk. The coefficient of this term is $-c/3$, as expected.  

As we approach $L_0$ we find
\eqn\sssaav{S = {c\over 3}\left[\sqrt{\alpha x_\infty} +{\pi\over 4}{1\over \alpha^{1/2}} - {\pi\over 32}{1\over \alpha^{3/2}} + {\cal O}\left({1\over \alpha^{5/2}}\right) \right], \quad \alpha = \lambda U_0^2, \quad x_\infty = {U_\infty^2\over U_0^2},
}
which simplifies using \lennnnv\ to
\eqn\sssaa{S = {c\over 3}\left[ {L_0\over L_{\Lambda}} + {\pi\over 2}\left({L\over L_0} - 1\right)^{{1\over 2}} - {\pi\over 16}\left({L \over L_0} - 1\right)^{{3\over 2}} + {\cal O}\left(\left({L\over L_0} - 1\right)^{5\over 2}\right) \right].
}
We note that there is no logarithmically divergent term. The entanglement entropy shows a square root area law correction at next--to--leading order. In local and Lorentz--invariant even--dimensional quantum field theories, however, in general the presence of a logarithmically divergent term is generic, and its coefficient is expected to be universal \CasiniHU.
 
\subsec{Case $\Psi > 0: \epsilon = 0$}

This case is studied in \ChakrabortyKPR. Setting $\alpha = \alpha_-$ in \sstrlen\ and \entros\ we find for the length $L$ and entanglement entropy $S$
\eqn\sstrlenn{{L \over 2\sqrt{\alpha'\lambda}}= \sqrt{1 + \alpha \over \alpha} E\left(\arcsin\sqrt{{1 + \alpha \over 1 + 2\alpha}}, \sqrt{1 + 2\alpha \over 1 + 2\alpha + \alpha^2}\right),
}
\eqn\entrosss{ \eqalign{S & = {\sqrt{k\alpha'}\over  2G_{\rm N}^{(3)}} {1\over \alpha + 1}\left\{ \left(2{ \alpha}  - \alpha^2{d\over d\xi} \right)\right.\cr
&\left.\left[{1 \over \xi + 1} \cdot \Pi \left(\arcsin\sqrt{{\alpha + 1\over 2\alpha + 1}\cdot \left( {1-{1\over x_\infty}}\right)}, {2\alpha + 1 \over (\xi + 1)( \alpha+ 1)}, \sqrt{{2\alpha+ 1 \over\alpha^2 + 2\alpha + 1}}\right)\right]_{\xi = 0} \right.\cr 
& \left. + \ F\left(\arcsin\sqrt{{\alpha + 1\over 2\alpha + 1}\cdot \left(1 - {1\over x_\infty}\right) }, \sqrt{{2\alpha + 1 \over\alpha^2 + 2\alpha + 1}}\right)\right\}, \quad \alpha = \lambda U_0^2, \quad x_\infty = {U_\infty^2\over U_0^2}.
}}

We find that in the large $U_0$ limit the interval length $L$ asymptotes to a minimum value which we denote by $L_0$ (this should cause no ambiguity)
\eqn\minl{L_0 = {\pi\sqrt{\alpha'\lambda}\over 2} = {1\over 2}\sqrt{\pi t}.
}
We note that there is a factor of 2 difference between \smalll\ and \minl. We find from \sstrlenn\ that the interval length $L$ has the following large $U_0$ expansion
\eqn\lexpans{{L\over L_0} = 1+ {2\over  \pi\alpha} + {3\pi - 16\over 16\pi \alpha^2}+ {\cal O}\left( {1\over \alpha^3}\right) , \quad \alpha = \lambda U_0^2.
}
Inverting the above equation one finds
\eqn\lexpansxx{\alpha = {2\over \pi \xi} + {3\pi - 16\over 32} + {\cal O}(\xi), \quad \xi = {L\over L_0} - 1.
}

The small $U_0$ expansion takes the form 
\eqn\llns{{L\over\sqrt{\alpha'}} = {2\over U_0}\left[1 - {\alpha^2\over 4}{\rm Log}(\alpha) + {3\over 8}\alpha^3 {\rm Log}(\alpha) + {\cal O}\left(\alpha^4\right) \right] , \quad \alpha = \lambda U_0^2.
}
We note that for a long interval the surface is deep inside the bulk in the $AdS_3$ region. We also note that the term linear in $\alpha$ is zero. Therefore, the correction starts, in this case, at order $\lambda^2$.

Inverting the above equation we find
\eqn\llnsxs{\alpha = \left({4\over \pi} {L_0\over L}\right)^2\left[1 - \left({4\over \pi} {L_0\over L}\right)^4{\rm Log}\left({4\over \pi} {L_0\over L}\right) + {\cal O}\left(\left({4\over \pi} {L_0\over L}\right)^6\right)\right].
}

In the large $L$ limit we find that the entanglement entropy $S$ has the following series expansion 
\eqn\entrlar{S = {c\over 3}\left[ {1\over 2}\alpha x_\infty  - {1\over 2}{\rm Log}(\alpha) - {1\over 4}\alpha + {\cal O}\left(\alpha^2{\rm Log}(\alpha)\right)\right],\quad \alpha = \lambda U_0^2, \quad x_\infty = {U_\infty^2\over U_0^2},
}
which using \llnsxs\ gives
\eqn\entrlar{S = {c\over 3}\left[ {1\over 2}\cdot {L_0^2\over L^2_{\Lambda}}\cdot {16\over \pi^2} - {\rm Log}\left({L_0\over L}\cdot {4\over \pi}\right) - {1\over 4}\left({L_0^2\over L^2}\cdot {16\over \pi^2}\right) +  {\cal O}\left({L_0^2\over L^2}\cdot {16\over \pi^2}\right)^2\right], \quad U_\infty := {2\sqrt{\alpha'}\over L_\Lambda}.
}
The (leading) logarithmic term is due to the deep $AdS_3$ region in the bulk. The coefficient of this term is $-c/3$, as expected. 

 As we approach $L_0$ the entropy $S$ takes the form 
\eqn\sss{S = {c\over 3}\left[{1\over 2}\alpha x_\infty - {1\over 2}{\rm Log}(\alpha) + {\cal O}\left({1\over \alpha}\right)\right],\quad \alpha = \lambda U_0^2, \quad x_\infty = {U_\infty^2\over U_0^2}.
}
Using \lexpansxx\ this gives
\eqn\sss{S = {c\over 3}\left[{1\over 2}\cdot {16\over \pi^2}\cdot{L^2_0\over L^2_{\rm \Lambda}}+ {1\over 2}{\rm Log} \left( {\pi\over 2 }\left({L\over L_0} - 1\right)\right) + {\cal O}\left({L\over L_0} - 1\right)\right].
}
In this limit the geometry is a linearly varying dilaton background. We note that in this case we have a logarithmically divergent term as opposed to the former $\Psi = 0$ case. However, the coefficient of this term is $c/6$. 

\subsec{Case $\Psi > 0: \epsilon \neq 0$}
In the large $U_0$ limit we find from \sstrlen\  that the interval length approach a minimum value $L_0$
\eqn\limL{L_0 = {\pi \sqrt{\alpha'\lambda}\over 2} = {1\over 2}\sqrt{\pi t},
}
that is determined solely by $\lambda$. 

The length $L$ has the following large $U_0$ expansion
\eqn\lexpuiiiu{{L\over L_0} = 1 + {2 - \delta^2\over \pi (1 - \delta^2)\alpha} + {3\pi - 16 + 2(4 - \pi)\delta^2 - \pi \delta^4\over 16 \pi (1 - \delta^2)^2\alpha^2} + {\cal O}\left({1\over \alpha^3}\right), \quad \delta^2 = {\epsilon^2\over \lambda}, \quad \alpha = \lambda U_0^2.
}
Inverting this we find 
\eqn\lexpuhhiiiu{\alpha = {2 - \delta^2\over \pi(1 - \delta^2)\xi} + {3\pi - 16 +2(4 - \pi)\delta^2 - \pi \delta^4\over 32 - 48\delta^2 + 16\delta^4}+{\cal O}(\xi), \quad \xi = {L\over L_0} - 1.
}

The small $U_0$ expansion takes the form
\eqn\llnsuuuds{{L\over\sqrt{\alpha'}} = {2\over U_0}\left[1 -{\delta^2\over 4}\alpha {\rm Log}(\alpha) - {\alpha^2\over 4}(1 + \delta^2 p){\rm Log}(\alpha) + {\cal O}(\alpha^3)\right], \quad \delta^2 = {\epsilon^2\over \lambda}, \quad \alpha = \lambda U_0^2,
}
here $p$ is a polynomial in $\delta^2$. The leading term corresponds to the deep $AdS_3$ region. We note that in this case the correction starts at order $\epsilon^2$.

Inverting the above equation one finds
\eqn\llnsvvvds{\alpha = \left({4\over \pi}{L_0\over L}\right)^2\left(1 - \delta^2 \left({4\over \pi}{L_0\over L}\right)^2 {\rm Log}\left({4\over \pi}{L_0\over L}\right) - \left({4\over \pi}{L_0\over L}\right)^4(1 + \delta^2 p){\rm Log}\left({4\over \pi}{L_0\over L}\right) + {\cal O}\left(\left({4\over \pi}{L_0\over L}\right)^6\right)\right).
}

In the large $L$ limit we find the following expansion for the entropy $S$
\eqn\llnsvvvdststss{S = {c\over 3}\left[{1\over 2}\alpha x_\infty \sqrt{1 - \delta^2} -{1\over 2}{\rm Log}(\alpha) - {1\over 4}\alpha\left({1\over (1 - \delta^2)^2} + {1\over 2}\delta^2 {\rm Log}(\alpha)\right) + {\cal O}\left(\alpha^2\right)\right],
}
which upon using \llnsvvvds\ gives
\eqn\llnsvvvddsststss{S = {c\over 3}\left[\left({4L_0\over \pi L_\Lambda}\right)^2 \sqrt{1 - \delta^2\over 4} -{\rm Log}\left({4L_0\over \pi L} \right) -{1\over 4}  \left({4L_0\over \pi L} \right)^2\left[{1\over (1 - \delta^2)^2} - \delta^2 {\rm Log}\left({4L_0\over \pi L} \right) \right]+{\cal O} \left({L_0\over L} \right)^4\right],
}
where $L_\Lambda$ is defined in \entrlar. The leading logarithmic term as in the previous cases corresponds to the $AdS_3$ region found deep inside the bulk. The coefficient of this term is $-c/3$, as expected. 

As we approach $L_0$ we find 
\eqn\ssds{S = {c\over 3}\left[{1\over 2}\alpha x_\infty - {2-\delta^2\over 4\sqrt{1 - \delta^2}}{\rm Log}(\alpha) + {\cal O}\left({1\over \alpha}\right)\right],
}
which using \lexpuhhiiiu\ gives 
\eqn\tesss{S = {c\over 3}\left[{1\over 2}\cdot {16\over \pi^2}\cdot{L^2_0\over L^2_{\Lambda}} \cdot\sqrt{1 - \delta^2} +{2-\delta^2\over 4\sqrt{1 - \delta^2}}{\rm Log} \left(\left({L\over L_0} - 1\right)\cdot {\pi(1 - \delta^2)\over 2 - \delta^2}\right) + {\cal O}\left({L\over L_0} - 1\right)\right].
}

We note that the leading logarithmic term has a coefficient that depends on $\delta^2$. 

In the next section we study in the above cases the Casin--Huerta entropic $c$--function.

\newsec{Casin--Huerta entropic $c$--function}

In quantum field theory entanglement entropy $S$ is ultraviolet divergent. It requires an ultraviolet cutoff to regularize the divergence. However, in two--dimensional local and Lorentz--invariant quantum field theories it is shown that the Casin--Huerta entropic $c$--function \CasiniES\ which is defined as
\eqn\cfun{C := L{\partial S\over \partial L},
}
is finite, and at fixed points of renormalization group flow it is proportional to the corresponding central charges. It is also a monotonic function of the interval length $L$ and independent of the ultraviolet cutoff that is introduced to regularize the entanglement entropy. The interval length $L$ is interpreted as a renormalization group scale. 

In this section we study the entropic $c$--function for the different cases we studied in the former section. We study in each case its monotonicity as a function of the renormalization group scale and its independence of the ultraviolet regulator.

\subsec{Case $\Psi = 0: \lambda = 0, \ \epsilon = 0$}

In this case we have
\eqn\cfuad{C = {c\over 3}.
}
This is the result for a two--dimensional holographic conformal field theory with central charge $c$ \CasiniES. It is independent of the interval length $L$. The entropic $c$--function is non--negative and constant.

\subsec{Case $\Psi = 0: \lambda = \epsilon^2 \neq 0$}

In this case we find that the entropic $c$--function $C(\alpha)$ is given by
\eqn\cfunc{C = {c\over 3}\sqrt{1 + \alpha}E\left(\sqrt{1\over 1+\alpha}\right), \quad \alpha = \lambda U_0^2.
}
We study the small and large $U_0$ limit, or equivalently the large and small $L$ limits of the entropic $c$--function \cfunc.  

In the large $L$ limit we find 
\eqn\cfct{C = {c\over 3}\left(1 + {2\over \pi^2 \xi^2}{\rm Log}\left({2\over \pi}\cdot \xi\right) + {\cal O}\left({1\over\xi^4}\right)\right), \quad L :=  \xi L_0, \quad L_0 = \pi \sqrt{\alpha'\lambda} = \sqrt{\pi t}.
}
In the small $L$ limit we find 
\eqn\cfffct{C = {c\over 3}\left( {\pi\over 4}\cdot {1\over \xi^{1\over 2}} + {5\pi\over 32}\cdot \xi ^{1\over 2}+ {\cal O}(\xi^{3/2})\right), \quad \xi = {L\over L_0} - 1.
}

One can think of $L$ as a renormalization group scale. We note that the entropic $c$--function increases as we ascend the renormalization group, and it diverges in the ultraviolet at $L_0$. At short distances it diverges as
\eqn\cdiv{C \sim \xi^{-{1\over 2}}, \quad \xi = {L\over L_0} - 1.
}
This is the case since in the ultraviolet the theory is not governed by a fixed point. We also note that the entropic c--function is independent of the ultraviolet cutoff that we introduced to regularize the entanglement entropy. The variable $\alpha$ can be expressed using the result \sstrlenss\ in terms of the interval length $L$ and the coupling $t$ to write \cfunc\ as a function of only $L$ and $t$.

\subsec{Case $\Psi > 0: \epsilon = 0$}
This case is studied in \ChakrabortyKPR\foot{It is also studied in closely related works \refs{\LewkowyczXSE, \ \GrieningerZTS}.}. In this case we find that the entropic $c$--function $C(\alpha)$ in closed--form is given by
\eqn\cfuncc{C = {c\over 3}(1 + \alpha)E\left(\arcsin\sqrt{1 + \alpha\over 1 + 2\alpha},\sqrt{1 + 2\alpha \over (1+\alpha)^2}\right), \quad \alpha = \lambda U_0^2.
}

Using the results \lexpansxx\ and \llnsxs\ for $U_0$ we find in the large $L$ limit
\eqn\ccfuncth{C  =  {c\over 3}\left(1 + {8\over \pi^2 \xi^2} + {\cal O}\left({1\over \pi \xi}\right)^4\right), \quad \xi := {L\over L_0}, \quad L_0 = {\pi\sqrt{\alpha' \lambda}\over 2} = {1\over 2}\sqrt{\pi t},
}
and in the small $L$ limit
\eqn\ccfthr{C  = {c\over 3}\left({1\over 2}\left(1 - {1\over \xi}\right)^{-1} + {\cal O}\left(1 - {1\over \xi}\right)\right), \quad \xi := {L\over L_0}.
}

We also note in this case that the entropic $c$--function is non--negative and increasing. At short distances it diverges as
\eqn\cddive{C \sim \xi^{-1}, \quad \xi = {L\over L_0} - 1.
}

Our results \ccfuncth\ and \ccfthr\ are in agreement with the corresponding analyses in \ChakrabortyKPR. Our result \cfuncc\ gives a non--perturbative answer and it can be written as a function of the interval length $L$ and the coupling $t$. We also note from \cfuncc\ that the entropic c--function is independent of the ultraviolet cutoff.

\subsec{Case $\Psi > 0: \epsilon \neq 0$}

In this case we find that the $c$--function $C(\alpha, \chi)$ is given by
\eqn\cfunge{C = {c\over 3}\sqrt{(1 + \alpha)(1 + \alpha\chi)}E\left(\arcsin\sqrt{1 + \alpha\chi\over 1 + 2 \alpha \chi}, \sqrt{1 + 2\alpha\chi\over (1 + \alpha)(1 + \alpha\chi)}\right), 
}
where 
\eqn\varDEDEDE{\alpha = \lambda U_0^2, \quad \Psi = \lambda \chi.
}

In this case the large $L$ limit of the entropic $c$--function takes the form
\eqn\cdcf{C  =  {c\over 3}\left(1 + {8\over \pi^2 \xi^2}\left({1\over{\chi^2}} + (1 - \chi ){\rm Log}\left({\pi \xi\over 4}\right)\right) + {\cal O}\left({1\over \pi \xi}\right)^4\right), 
}
where
\eqn\cdcfSO{\xi := {L\over L_0}, \quad L_0 = {\pi\sqrt{\alpha' \lambda}\over 2} = {1\over 2}\sqrt{\pi t}.
}

In the small $L$ limit we find
\eqn\ccf{C  = {c\over 3}\left({1+ \chi\over4\sqrt{\chi}}\left(1 - {1\over \xi}\right)^{-1} + {\cal O}\left(1 - {1\over \xi}\right)\right), \quad \xi := {L\over L_0}.
}

In this case also the entropic $c$--function is non--negative, ultraviolet cutoff independent and increasing. In the ultraviolet it diverges as
\eqn\cfundiv{C \sim \chi^{-{1\over 2}}\cdot \xi^{-1}, \quad \xi = {L\over L_0} - 1.
}

We note that setting $\chi = 1$ in \cfunge\ gives \cfuncc, and setting $\chi = 0$ gives \cfunc. At $\alpha = 0$ it gives \cfuad. The entropic $c$--function $C$ \cfunge\ and the interval length $L$ \sstrlen\ satisfy the curious relation 
\eqn\clenrexn{{C\over c_0} = {L\over l_0}\cdot \sqrt{1 + \alpha\chi}, \quad l_0 = {2\sqrt{\alpha'}\over U_0}, \quad c_0 = {c\over 3}, \quad \alpha = \lambda U_0^2,  
}
where $l_0$ and $c_0$ can be thought of as the interval length $L$ and the entropic $c$--function $C$ at $\lambda = 0$ or in the infrared, respectively.

\newsec{Discussion}

In this paper we computed the (Von Neumann) entanglement entropy and the entropic $c$--function for an interval of length $L$. We found that the entropic $c$--function is given by
\eqn\cfungeef{C = {c\over 3}\sqrt{(1 + \alpha)(1 + \alpha\chi)}E\left(\arcsin\sqrt{1 + \alpha\chi\over 1 + 2 \alpha \chi}, \sqrt{1 + 2\alpha\chi\over (1 + \alpha)(1 + \alpha\chi)}\right), 
}
where\foot{In terms of the field theory side deformation parameters $\delta^2$ is given by ${\pi \mu^2\over 8 t}$, and $\mu = 2\mu_+ = 2\mu_-$. }
\eqn\varFFFF{\alpha = \lambda U_0^2, \quad \chi = 1 - \delta^2, \quad \delta^2 = {\epsilon^2\over \lambda} = {\pi \mu^2\over 8 t}.
}
The variable $\alpha$ is related to the interval length $L$ via
\eqn\sstrlennHHHH{{L \over 2\sqrt{\alpha'\lambda}}= \sqrt{1 + \alpha \over \alpha} E\left(\arcsin\sqrt{{1 + \alpha\chi \over 1 + 2\alpha\chi}}, \sqrt{1 + 2\alpha\chi \over (1 + \alpha)(1 + \alpha\chi)}\right), \quad \alpha'\lambda = {t\over \pi}.
}

We found that the entropic c--function is non--negative and ultraviolet cutoff independent. This is required for a theory that is internally consistent and it is a non--trivial consistency check. Therefore, this provides further evidence that the deformed theory is very sound and under control. We also found that along the renormalization group upflow towards the ultraviolet it is non--decreasing. At long distances it is proportional to the central charge of the original conformal field theory. At short distances it diverges. This is the case since in the ultraviolet the deformed theory is not governed by an ultraviolet fixed point. The minimum distance at which the entropic c--function diverges sets the nonlocality scale of the theory. 

In \figc\ we show plots of the entropic c--function $C$ \cfungeef\ as a function of the interval length $L$ \sstrlennHHHH\ for $\delta = 0$ and $\delta^2 = 1$. In the plots we normalized $C$ with respect to $c_0 = {c\over 3}$ and $L$ with respect to $L_0 = {1\over 2}\sqrt{\pi t}$.

In the case in which $F > 0$, the entropic $c$--function diverges in the ultraviolet as
\eqn\cfundivv{C \sim \chi^{-{1\over 2}}\cdot \xi^{-1}, \quad \xi = {L\over L_0} - 1, \quad L_0 := {\pi\sqrt{\alpha'\lambda}\over 2} = {1\over 2}\sqrt{\pi t}.
}
 Note that $L_0$ is determined solely by $\lambda$; it is independent of $\epsilon$.

In the case in which $F = 0$, the entropic $c$--function diverges at short distances as
\eqn\cdivvv{C \sim \xi^{-{1\over 2}}, \quad \xi = {L\over L_0} - 1, \quad L_0 := \pi \sqrt{\alpha'\lambda} = \sqrt{\pi t}.
}
Note that the minimum interval length $L_0$ in this case is twice larger than the corresponding length in the former case. This is depicted in \figc\ with $\delta = 0$ and $\delta^2 = 1$.

We note that the exponent of $\xi$ in the case in which $F > 0$ is $-1$, and in the case in which $F = 0$ it is $-{1/2}$. This is due to the fact that at short distances the entanglement entropy for the case in which $F > 0$ \tesss\ contains a logarithmically divergent term (with a coefficient that depends on $\chi$) whereas in the case in which $F = 0$ \sssaa\ it does not contain a logarithmically divergent term. In the latter case the entanglement entropy area law exhibits a square root correction at next--to--leading order at short distances.
 
 It would be interesting to compute the entropic $c$--function directly in $J{\bar T}, \ T{\bar J}, \ T{\bar T}$ deformed field theory, for example in perturbation theory, and compare it with the bulk calculation result \cfungeef\ (where $\alpha$ is given by \sstrlennHHHH). It would be also nice to understand better the theory corresponding to the case in which $F = 0$. We leave these for future work.
 
 In this paper we mainly focused on the case in which $F \geq 0$. As we mentioned in the introduction, in the case $F < 0$ or equivalently $\Psi < 0$, however, the bulk geometry is singular and/or it has either closed timelike curves or no timelike direction. If we look, for example, the case where $\epsilon = 0$, the signature of the boundary (spacetime) metric changes signs as we turn on the coupling $\lambda$. The boundary (spacetime) spatial coordinate $x$ becomes temporal at the outset of the deformation and it is not clear how to consistently apply the holographic prescription in this setting. That is, it is not clear as to whether or not we should consider the region beyond the finite radial distance at which the bulk metric signature changes signs. The holographic prescription may also have to be modified if we excise the region. In the field theory side we have states with complex energies. It would be nice to understand the entropic $c$--function in this case and compare it with results obtained from the closely related double trace deformations. For example, in the holographic proposal with radial cutoff the entropic $c$--function for an interval of length $L$ in the large $c$ limit is shown to be given by \LewkowyczXSE\foot{See also \GrieningerZTS\ for a related result.}  
 \eqn\cfunff{C = {c\over 3}\cdot {1\over \sqrt{1 - {2\pi \over 3}\cdot {ct\over L^2}}},
 }
 with $ct/L^2$ finite. It is also shown in \LewkowyczXSE\ that this result agrees with a field theory calculation performed in a $T{\bar T}$ deformed conformal field theory. We also left out the case $F \geq 0$ and $\mu_+ \neq \mu_-$ which requires using the Hubeny, Rangamani, and Takayanagi prescription \HubenyXT. We leave these for future work.
 
\bigskip\bigskip

\noindent{\bf Acknowledgements:} I thank D. Kutasov and S. Chakraborty for useful discussions and comments. This work is supported in part by DOE grant DE-SC0009924.

\appendix{A}{}

We collect the intermediate results required in section 3 to compute the interval length and the Von Neumann entanglement entropy, see \BF.

We have (with $u > a > b > c > d$),
\eqn\inttwo{\int_a^u dx{ \sqrt{x - c\over (x - a)(x - b)(x - d)}} = {2\over \sqrt{(a - c)(b - d)}}\left[(b - c)F(\varphi, k) + (a - b)\Pi(\varphi, n, k)\right],
}
\eqn\intthree{\int_a^u dx{ \sqrt{ x - b\over(x - a)(x - c)(x - d)}} = {2(a - b)\over \sqrt{(a - c)(b - d)}}\Pi(\varphi, n, k),
}
\eqn\intone{\int_a^u {dx\over x - b}\sqrt{x - c\over (x - b)(x - a)(x - d)} = {2\over a - b}\sqrt{a - c\over b - d}E(\varphi, k),
}
here
\eqn\vartwo{\varphi = \arcsin\sqrt{(b - d)(u - a)\over (a - d)(u - b)}, \quad n = {a - d\over b - d},\quad k = \sqrt{(b - c)(a - d)\over (a - c)(b - d)}.
}
We have (with $a > u \ge b > c, \ r \neq a$),
\eqn\inttwo{\int^a_u {dx\over x - r}{1\over \sqrt{ (a - x)(x - b)(x - c)}} = {2\over (a - r)\sqrt{a - c}}\Pi(\varphi, n, k),
}
here
\eqn\vartwo{\varphi = \arcsin\sqrt{a - u\over a - b}, \quad n = {a - b\over a - r},\quad k = \sqrt{a - b\over a - c}.
}

\listrefs
\bye
\end